\documentclass[twocolumn,twoside,slac_two,floatfix]{revtex4}
\usepackage{graphicx}
\usepackage{fancyhdr}
\usepackage{hyperref}
\begin{document}

\title{The CMS Computing System:  Successes and Challenges}

%

\author{Kenneth Bloom}
\affiliation{Department of Physics and Astronomy, University of Nebraska-Lincoln, Lincoln, NE 68588-0111, USA}

\begin{abstract}
  Each LHC experiment will produce datasets with sizes of order one
  petabyte per year. All of this data must be stored, processed,
  transferred, simulated and analyzed, which requires a computing
  system of a larger scale than ever mounted for any particle physics
  experiment, and possibly for any enterprise in the world. I discuss
  how CMS has chosen to address these challenges, focusing on recent
  tests of the system that demonstrate the experiment's readiness for
  producing physics results with the first LHC data.
\end{abstract}

\maketitle

\thispagestyle{fancy}


\section{The Problem}
Experiments at the Large Hadron Collider (LHC)~\cite{ref:LHC} will
produce tremendous amounts of data.  With instantaneous luminosities
of $10^{34}$~cm$^{-2}$s$^{-1}$ and a crossing rate of 40~MHz, the
collision rate will be about $10^9$~Hz.  But the rate for new physics
processes, after accounting for branching fractions and the like, is
of order $10^{-5}$~Hz, leading to the need to select events out of a
huge data sample at the level of $10^{-14}$.

What does this imply for the necessary scale of computing systems for
an LHC experiment, and for the Compact Muon Solenoid (CMS) in
particular?  The first run of the LHC in 2009-2010 is expected to be
quite long, with six million seconds of running time.  CMS plans to
record data at 300~Hz, leading to datasets of 2.2~billion events, once
dataset overlaps are accounted for.  Roughly as many events will be
simulated.  The size of the raw data from a single event is 1.5~MB
(and 2.0~MB for simulated data), already implying petabytes worth of
raw data alone from just the first year of operations.  All of this
data must be processed; detector data is reconstructed at a rate of
100~HS06-sec/event~\cite{ref:HS06} while simulated data is generated
and reconstructed at 1000~HS06-sec/event.  Given these parameters, the
CMS computing model~\cite{ref:CMSmodel} estimates that 400~kHS06 of
processing resources, 30~PB of disk and 38~PB of tape will be required
to handle just the first year of CMS data.

CMS has been developing a distributed computing model from the
very early days of the experiment. There are a variety of
motivating factors for this:  a single data center at CERN would
be expensive to build and operate, whereas smaller data centers
at multiple sites are less expensive and can leverage local 
resources (both financial and human).  But there are also many
challenges in making a distributed model work, some of which
are discussed here.

\begin{figure}[h]
\centering
\includegraphics[width=80mm]{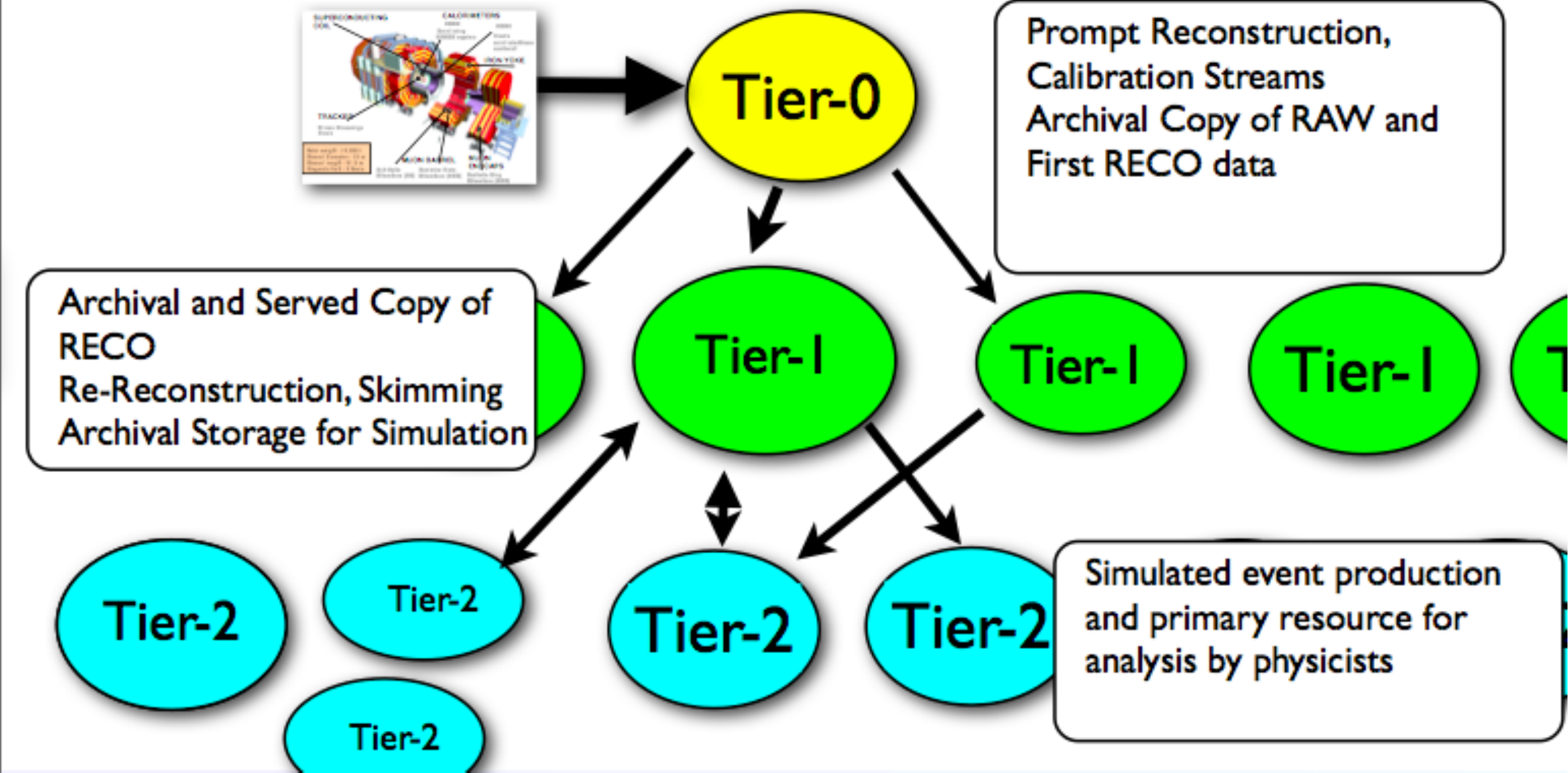}
\caption{Tiered hierarchy of the CMS distributed computing
  model.} \label{fig:tiers}
\end{figure}

The CMS distributed computing model~\cite{ref:CMSmodel} has different
computing centers arranged in a ``tiered'' hierarchy, as illustrated
in Figure~\ref{fig:tiers}, with experimental data typically flowing
from clusters at lower-numbered tiers to those at higher-numbered
tiers.  The different centers are configured to best perform their
individual tasks.  The Tier-0 facility at CERN is where prompt
reconstruction of data coming directly from the detector takes place;
where quick-turnaround calibration and alignment jobs are run; and
where an archival copy of the data is made.  The facility is typically
saturated by just those tasks.  There are seven Tier-1 centers in
seven nations (including at FNAL in the United States).  These centers
keep another archive copy of the data\footnote{If one would say that
  the data is not truly acquired until there are two safe copies of
  it, then the CMS data acquisition system stretches around the
  world.}, and are responsible for performing re-reconstruction of
older data with improved calibration and algorithms, and making skims
of primary datasets that are enriched in particular physics signals.
They also provide archival storage of simulated samples produced at
Tier-2.  There are about 40 Tier-2 sites around the world (including
seven in the U.S.); they are the primary resource for data analysis by
physicists, and also where all simulations done for the benefit of the
whole collaboration take place.  These centers thus host both
organized and chaotic computing activities.  (Tier-2 centers are
discussed further in Section~\ref{sec:t2descr}).

Of course, the Tevatron Run~II experiments have also created computing
systems of impressive scale.  But computing for CMS will be something
still different.  For instance, there will not be enough resources at
any single location to perform all analysis; CDF, by contrast, has
approximately equal resources at FNAL for reconstruction and analysis.
CMS in fact depends on large-scale dataset distribution away from CERN
for successful analysis computing.  At CMS, all re-processing resources
will be remote.  It is true that D0 does much of its re-processing off
the FNAL site, but this was put into place after other elements of the
computing system were commissioned.  Most notably, the commissioning
of the distributed computing model will be simultaneous with the
commissioning of the CMS detector, not to mention the search for new
physics that is the object of the experiment.  Given the stresses that
the system will face early on, we must take all steps possible to make
sure that the system is ready before we have colliding beams.

\section{STEP 09}
Such a step is a recent exercise called the Scale Testing of the
Experimental Program (STEP).  This was a multi-virtual organization
(VO) exercise performed in the context of the Worldwide LHC Computing
Grid (WLCG)~\cite{ref:WLCG}.  The primary goal for the WLCG was to
make sure that all experiments could operate simultaneously on the
grid, and especially at sites that are shared amongst VO's.  All
of the LHC VO's agreed to do their tests in the first two weeks of
June 2009.

For CMS, STEP~09 was not an integrated challenge.  This way,
downstream parts of the system could be tested independently of the
performance of upstream pieces.  The factorization of the tests made
for a much less labor-intensive test, as CMS also needed to keep focus
on other preparations for data-taking, such as commissioning the
detector through cosmic-ray runs, during this time.  CMS thus focused
on the pieces of the distributed system that needed the greatest
testing, and had the greatest VO overlap.  These were data transfers
from tier to tier; the recording of data to tape at Tier~0; data
processing and pre-staging at Tier~1, and the use of analysis
resources at Tier~2.  The specific tests and their results are
described below.

\subsection{Data transfers}

Data transfer is a key element of the CMS computing model; remote
resources are of little use if data files cannot be transferred in and
out of them at sufficient rates for sites to be responsive to the
evolving demands of experimenters.  Several elements of data transfer
were tested in STEP~09.  Tier-1 sites must archive data to tape at
close to the rate that it emerges from the detector, if backups of
transfers and disk space are to be avoided.  In STEP~09, CMS exported
data from Tier~0 at the expected rates to the Tier-1 sites for
archiving.  Latencies were observed between the start of the transfer
and files being written to tape, and in some cases these latencies had
very long tails, with the last files in a block of files being written
very long after the first files were.  Latencies were correlated with
the state of the tape systems at the individual sites; they were
longer when there were known backlogs at a given site.

While each Tier-1 site only has custodial responsibility for a
particular fraction of the entire RECO-level sample, which contains
the full results of event reconstruction, every Tier-1 site hosts a
full copy of the analysis-object data (AOD) sample, which contains
only a summary of the reconstruction.  When a re-reconstruction pass
is performed at a particular Tier~1, new AOD's are produced for the
fraction of the data that the site has archived, and then those
particular AOD's must be distributed to the other six sites.  This
results in substantial traffic among the seven sites.  These transfers
were tested in STEP~09 by populating seven datasets at the seven
sites, with sizes proportional to the custodial fraction, and then
subscribing these datasets to the six other sites for transfer.  The
total size of the dataset was 50~TB, and the goal was to complete all
of the transfers in three days.  An aggregate sustained transfer rate
of 1215~MB/s was required to achieve that goal, and a rate of 989~MB/s
was achieved.

\begin{figure}[h]
\centering
\includegraphics[width=80mm]{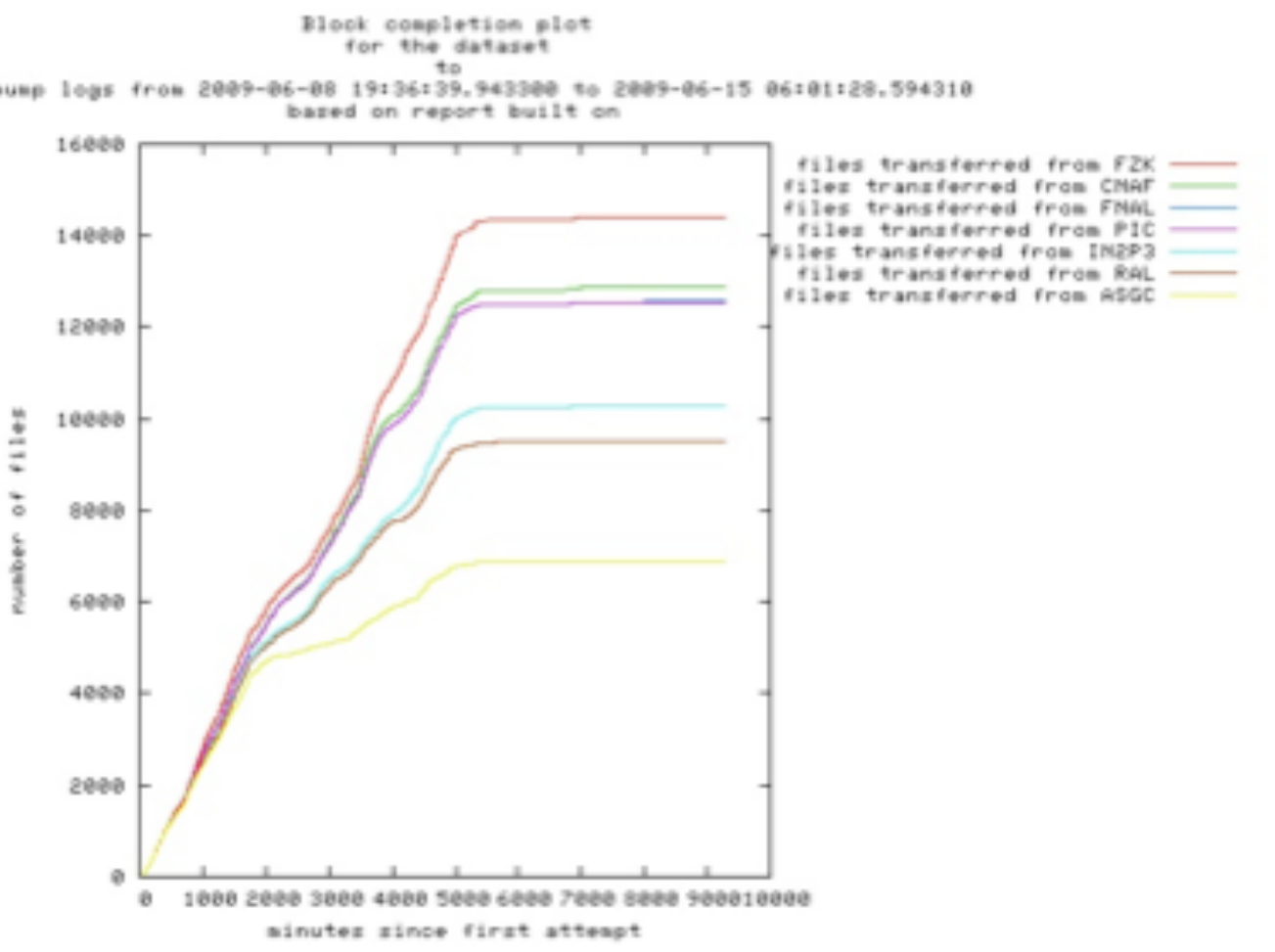}
\caption{Sources of file transfers versus time for a dataset that
  originated at ASGC Once files become available at other sites,
  transfers tend to originate from there rather than ASGC
  (yellow).} \label{fig:reroute}
\end{figure}

One interesting feature of these transfers was that it demonstrated
the re-routing capabilities of the PhEDEx transfer
system~\cite{ref:PhEDEx}.  PhEDEx attempts to route files over the
fastest links available.  If site A is the original source of a file
and both sites B and C wish to acquire them, then if B gets a file
from A before C does, and the network link between B and C is faster
than that between A and C, then site C will obtain the file from site
B rather than the originating site A.  This is illustrated in
Figure~\ref{fig:reroute}, which shows which sites serve as the sources
for files that were originally at ASGC in Taiwan, the Tier-1 site that
is furthest from all the others in CMS.  In the early stages of the
transfer, ASGC is the only source of the files.  But once the files
have started to arrive in Europe, other European Tier-1's start to get
the files from their nearest neighbors rather than ASGC.  In the end,
only about half of the transfers of the ASGC dataset actually
originated at ASGC.  CMS is learning how to best take advantage of
such behavior.

\begin{figure}[h]
\centering
\includegraphics[width=80mm]{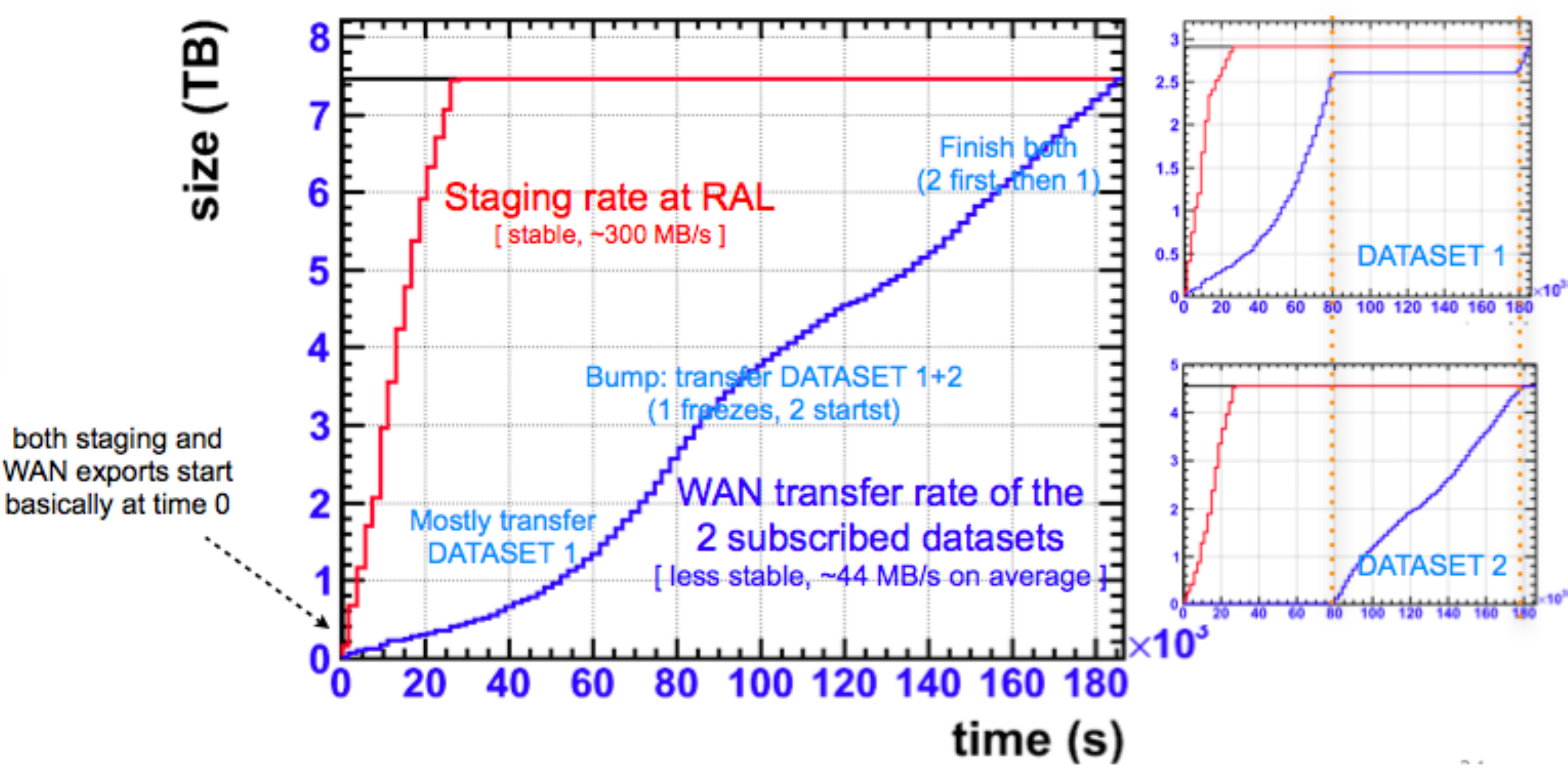}
\caption{Example of latency effects in two datasets being transferred
  from a Tier-1 site to a Tier-2 site.} \label{fig:t1t2}
\end{figure}

Finally, transfers from Tier-1 to Tier-2 are important for getting
data into the hands of physicists.  These transfers typically involve
pulling data off tape at the Tier-1 site so that disk-resident files
can then be copied to Tier-2 disk pools.  STEP~09 testing of these
transfers focused on stressing Tier-1 tape systems by transfering
files that were known not to be on disk at the originating sites.  In
general, the target transfer rates were achieved, with the expected
additional load on tape systems observed.  One interesting feature
that was observed is shown in Figure~\ref{fig:t1t2} for the case of
two datasets being transferred from the Tier-1 site at RAL in the UK
to a nearby Tier-2 site.  Both datasets were brought to disk pretty
quickly, and the first dataset was mostly transferred after that.
However, the transfer of that dataset was stalled for a while as the
second dataset was transferred in its entirety.  Since only complete
blocks of files are visible to CMS jobs, the first dataset was
probably not in a useable state while the second dataset was being
transferred.  CMS is studying techniques to avoid such issues.

\subsection{Tier~0}
The primary responsibility of the Tier-0 facility is to do a first
pass reconstruction of the raw data, and then to save an archival copy
of the raw data and the reconstruction output.  In STEP~09, the Tier-0
tape system CMS stressed by running I/O intensive jobs at the same
time that other experiments ran similar jobs.  Could CMS archive data
to tape at sufficient rates while other experiments were doing the
same?  ``Sufficient'' is hard to define, as the 50\% duty cycle of the
LHC allows time to catch up between fills.  CMS estimated that a
500~MB/s tape-writing rate would be sufficient.

\begin{figure}[h]
\centering
\includegraphics[width=80mm]{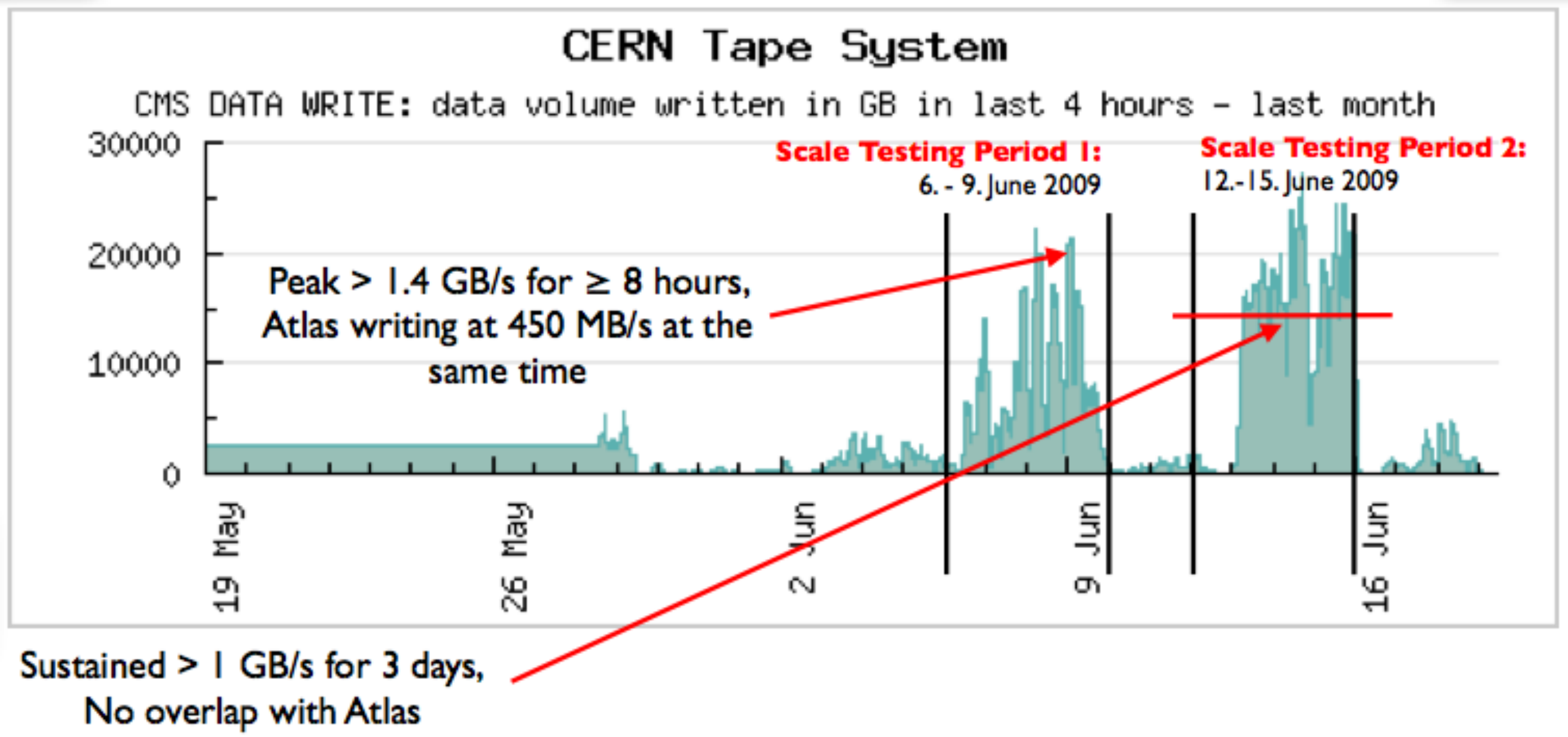}
\caption{Tier-0 tape-writing rates during STEP~09.  The target rate
for CMS was 500~MB/s or 7200~GB/4~hours.} \label{fig:t0write}
\end{figure}

The tape-writing test schedule was constrained by the need to handle
real detector data from cosmic-ray runs during the STEP~09 period,
leading to two test periods of four and five days.  The results are
shown in Figure~\ref{fig:t0write}.  In both periods, the target rate
was easily exceeded, even with ATLAS also writing at a high rate
during one of the periods.  The only problem that was encountered was
the limited amount of monitoring information for the Tier-0 facility.

\subsection{Tier~1}

The Tier-1 sites hold custodial copies of datasets, and will be
re-reconstructing those events multiple times.  In 2010, CMS expects
to do three re-processing passes that will take four months each.  In
the early stages of the experiment, when data sizes are small, all of
the raw data and several versions of the reconstruction will fit onto
disk pools at the Tier-1 sites, making for efficient processing.  But
as the collected dataset gets bigger, it will have to be staged from
tape to disk for re-processing.  This is potentially inefficient; one
wouldn't want to have re-processing jobs occupying batch slots and
waiting for file staging.  Thus some pre-staging scheme is required to
maximize CPU efficiency.  The pre-staging has never been tested by CMS
on this scale or with such coordination.  STEP~09 exercises at Tier~1
investigated the prestage rates and stability of the tape systems, and
the ability to perform rolling re-reconstruction.

A rolling re-processing scheme was established for the exercise.  On
Day 0, sites pre-staged an amount of data that could be
re-reconstructed in a single day from tape to disk.  On Day 1, that
data was processed while a new batch of data was pre-staged.  On Day
2, the Day 0 data was purged from disk, the Day 1 data was processed,
and new data was again pre-staged.  This was repeated throughout the
exercise period.  How much data was processed varied by the custodial
fraction at each site.  CMS does not yet have a uniform way of
handling pre-staging within the workload management system.  Three
different implementations emerged across the seven Tier-1 sites.  All
three worked, and the experienced gained will be used to design a
uniform pre-staging system for long-term use.

The target pre-staging rates for each site are given in
Table~\ref{tab:t1stage}.  Also shown are the best one-day average
rates that were achieved during the exercise.  As can be seen, all
sites were able to achieve the targets, although there were some
operational problems during the two weeks.  The FZK tape system was
unavailable at first, and the performance was not clear once it was
available.  IN2P3 had a scheduled downtime during the first week of
STEP~09.  The large rates required at FNAL triggered problems at first
that led to a backlog, but these were quickly solved.

\begin{table}
\begin{center}
\caption{Target and best achieved pre-staging rates at Tier-1 sites
during STEP~09.\label{tab:t1stage}}
\begin{tabular}{|l|c|c|}\hline
Site & Target (MB/s) & Best (MB/s)\\\hline
FZK & 85 & x \\
PIC & 50 & 142 \\
IN2P3 & 52 & 120 \\
CNAF & 56 & 380 \\
ASGC & 73 & 220 \\
RAL & 40 & 250 \\
FNAL & 242 & 400 \\\hline
\end{tabular}
\end{center}
\end{table}

\begin{figure}[h]
\centering
\includegraphics[width=80mm]{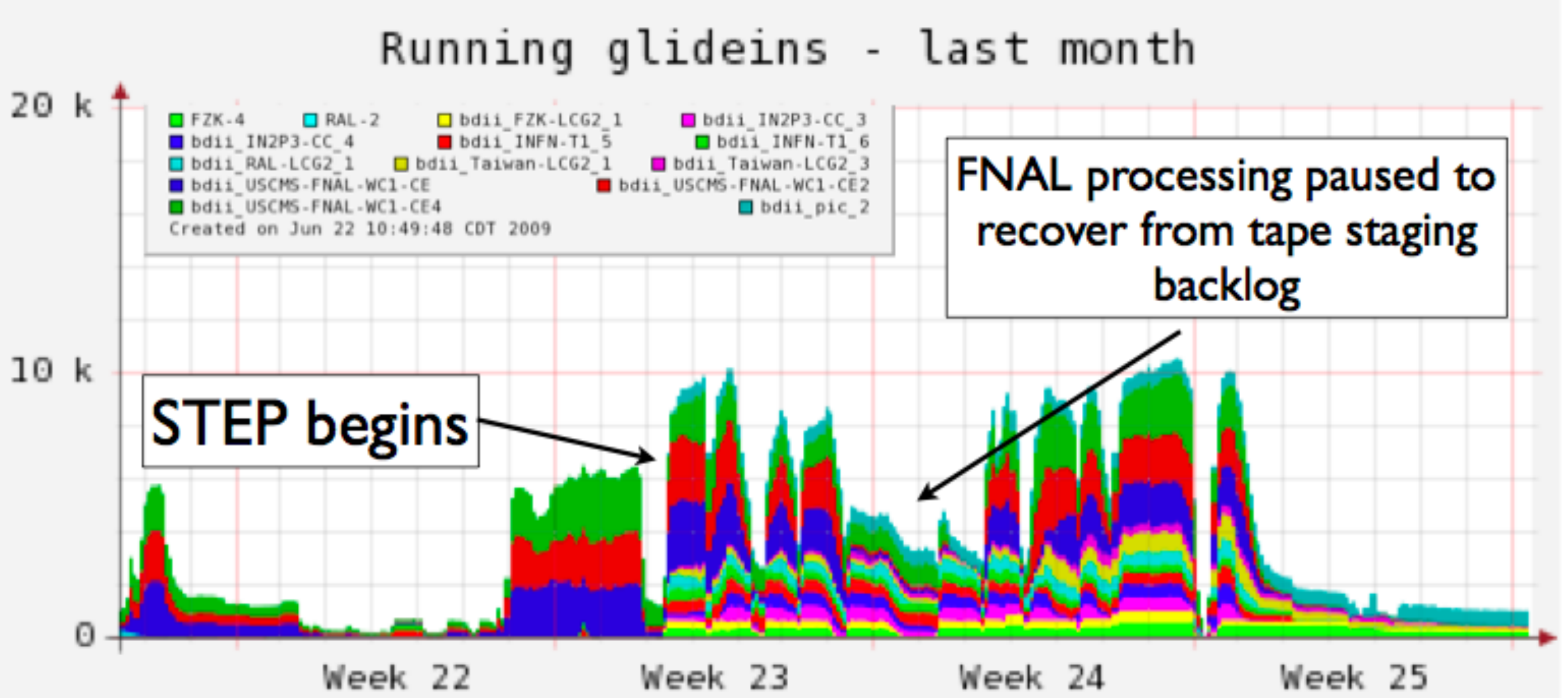}
\caption{Re-processing jobs at Tier-1 sites during STEP09.} \label{fig:t1proc}
\end{figure}

The re-processing operations ran quite smoothly.  A single operator
was able to submit many thousands of jobs per day using glide-in
pilots, as shown in Figure~\ref{fig:t1proc}.  (Note the effect of the
backlog at FNAL mentioned above.)  There was no difficulty in getting
the pledged number of batch slots from sites, and fair-share batch
systems appeared to give each experiment the appropriate resources.

\begin{figure}[h]
\centering
\includegraphics[width=80mm]{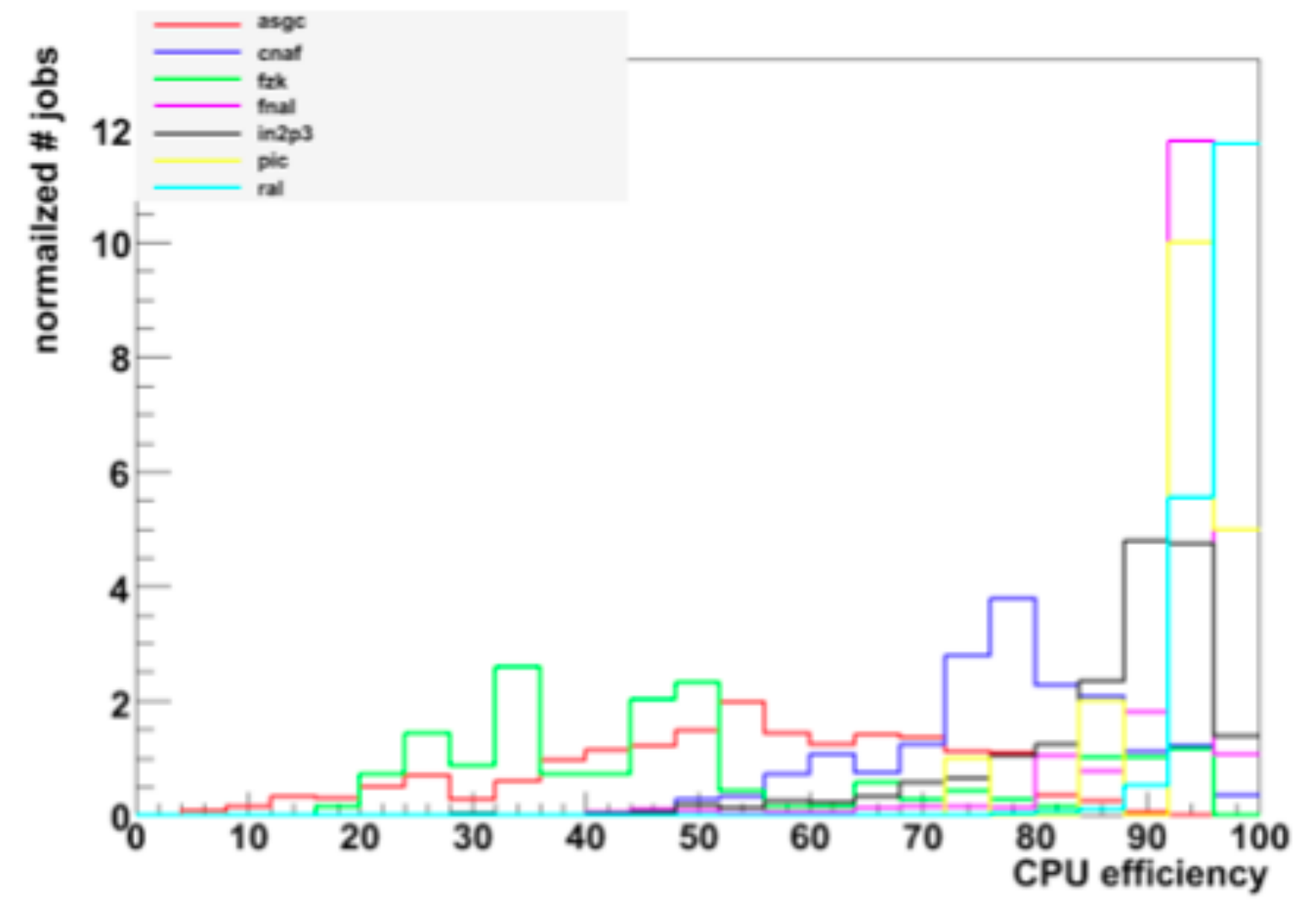}
\caption{CPU efficiency at Tier-1 sites during STEP09.} \label{fig:t1eff}
\end{figure}

The efficiency of the re-processing jobs is reflected in the ratio
of CPU time consumed by the jobs to the wall-clock time that the
job spends using a batch slot.  This ratio should be near one if
jobs are not waiting for files to come off tape.  Figure~\ref{fig:t1eff}
shows the efficiency for jobs on a typical STEP~09 day.  Efficiency 
varies greatly across sites, which bears more investigation.  However,
pre-staging, which was used here, is generally observed to greatly
improve the efficiency.

\subsection{Tier~2}
\subsubsection{Overview of CMS analysis model}
\label{sec:t2descr}
The CMS data analysis model depends greatly on the distribution of
data to Tier-2 sites and the subsequent submission of analysis jobs
to those sites.  We review those elements of the model here.

In CMS, analysis jobs go to the data, and not the other way around, so
it is important to distribute data for the most efficient use of
resources.  The nominal storage available at a Tier-2 site is 200~TB;
with about 40 functional Tier-2 sites, this is a huge amount of
storage that must be partitioned in a sensible way.  At each site, the
available disk space is managed by different parties ranging from the
central CMS data-operations group to large groups of users to
individual users, leading to a mix of central and chaotic control.  A
small amount of disk, about 10~TB, is set aside for as staging space
for centrally-controlled simulation production.  30~TB at each site is
designated as centrally-controlled; CMS will place datasets of wide
interest to the collaboration in this space.  Another 30-90~TB of
space, divided into 30~TB pieces, is allocated to individual physics
groups in CMS for distribution and hosting of datasets that are of
greatest interest to them.  There are 17 such groups in CMS.
Currently no site supports more than three groups and no group is
affiliated with more than five sites; the seven U.S. Tier-2 sites
support all 17 groups.  As a result, there are a manageable number of
communication channels between sites and groups, making it easier to
manage the data placement across the far-flung sites.  The remainder
of the space at a Tier-2 site is devoted to local activities, such as
making user-produced files grid accessible.

\begin{figure}[h]
\centering
\includegraphics[width=80mm]{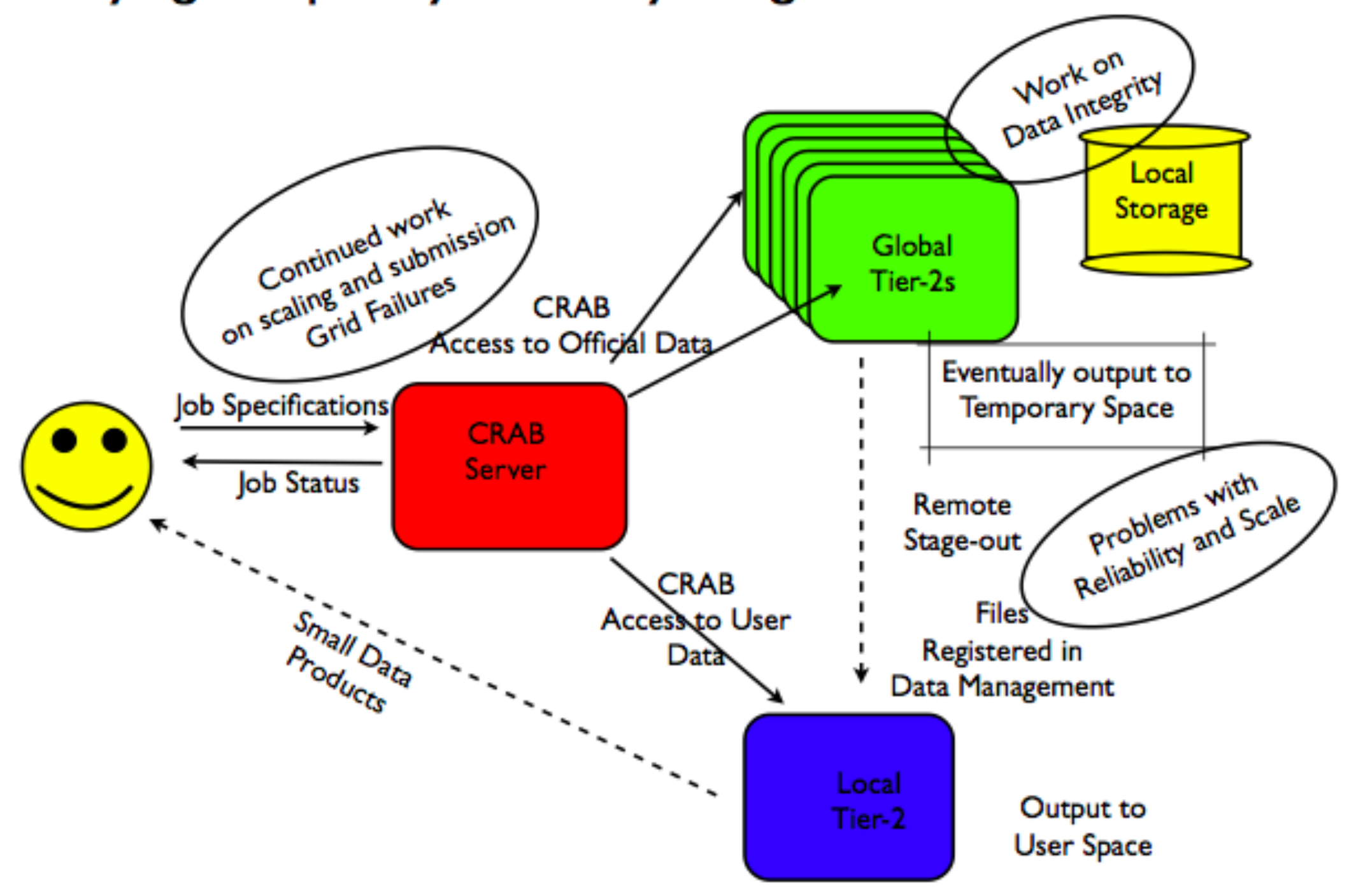}
\caption{Schematic diagram of user analysis job
  workflow.} \label{fig:workflow}
\end{figure}

CMS physicists must then be able to access this data.  All analysis
jobs are submitted over the grid.  To shield the ordinary user from
the underlying complexity of the grid, CMS has created the CMS Remote
Analysis Builder (CRAB)~\cite{ref:CRAB}.  A schematic diagram of how
the grid submission of an analysis job works is shown in
Figure~\ref{fig:workflow}.  A user creates a CRAB script that
specifies the job, including the target dataset and the analysis
program.  The user submits the script to a CRAB server with a one-line
command.  The server then determines where the dataset is located.
The dataset in question could be either an official CMS dataset, or
one created by a user that is resident at a Tier-2 site.  The job is
then submitted by the server to the appropriate site through the grid
for processing.  If the user is creating significant output, that
output can be staged to the user's local Tier-2 site, and the files
can be registered in the data management system for processing by a
future CRAB job.  Needless to say, many elements of the system must
succeed for the user to have a successful job.  Those of greatest
concern at the moment are the scaling of grid submissions, data
integrity at the Tier-2 sites, and reliability and scaling issues for
stageout of user output.

\subsubsection{STEP~09 tests}
\begin{figure}[h]
\centering
\includegraphics[width=80mm]{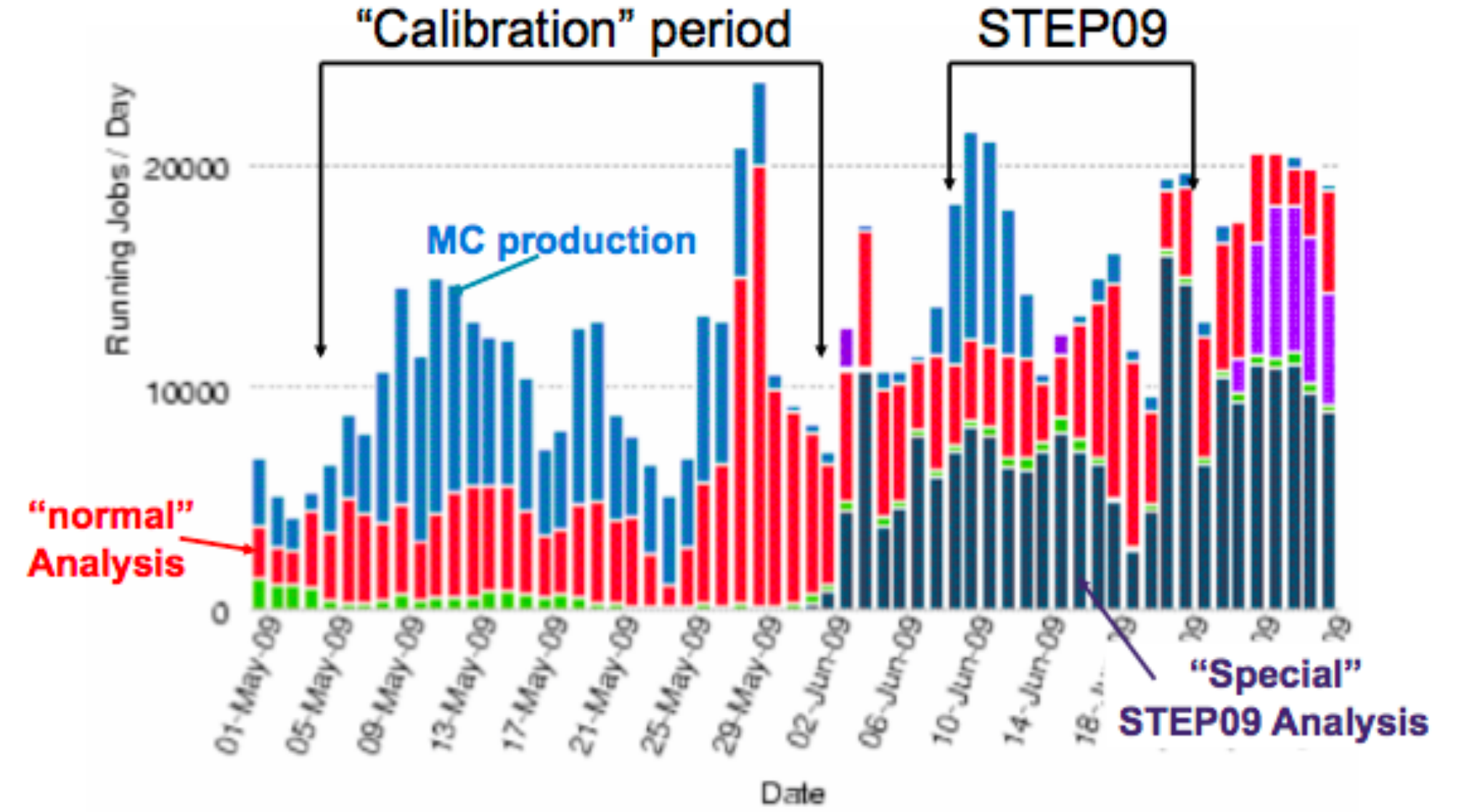}
\caption{Number of jobs running per day at Tier-2 sites before and
  during STEP~09.} \label{fig:t2slots}
\end{figure}

50\% of the pledged processing resources at Tier-2 sites are targeted
for user analysis.  At the moment, this is about 8,000 batch slots.
The primary goal of STEP~09 tests at Tier~2 was to actually fill that
many slots.  Figure~\ref{fig:t2slots} shows the number of running jobs
per day at the Tier-2 sites before and during STEP~09.  All types of
jobs that ran at the sites are indicated -- simulation production,
normal analysis run by users throughout CMS, and the extra analysis
jobs that were submitted for the exercise.  Between normal and STEP~09
analysis jobs, the pledged analysis resources were more than
saturated, with no operational problems at the sites.  This apparent
spare capacity suggests that CMS could be making better use of the
analysis resources.  Indeed, in the month before STEP~09, only five
out of 48 sites were devoting more than 70\% of their analysis
resources to analysis jobs.  During STEP~09, 33 sites did so.  This
bodes well for the onslaught of user jobs that we expect when LHC
data-taking begins.

\begin{figure}[h]
\centering
\includegraphics[width=80mm]{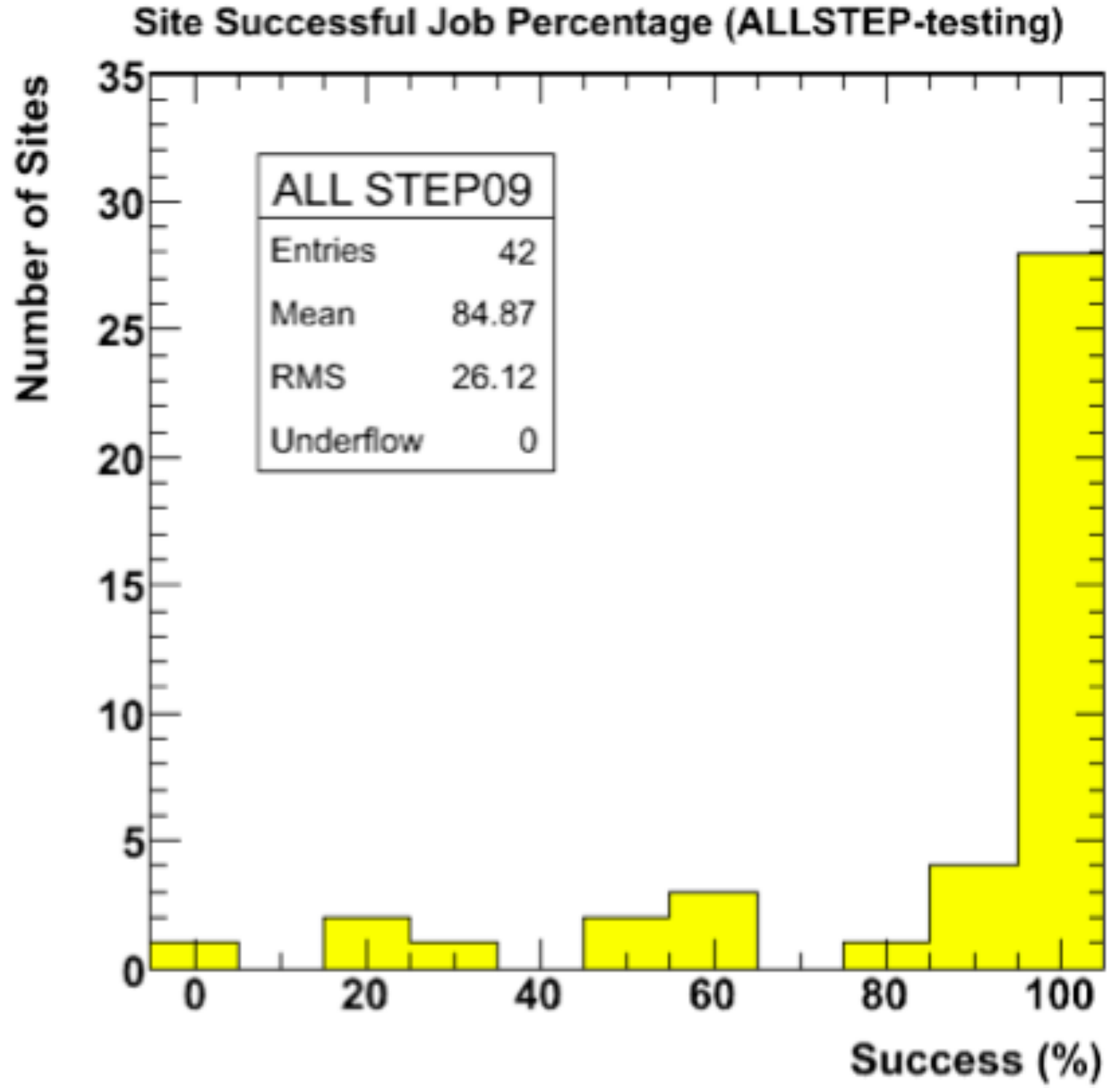}
\caption{Success rate of STEP~09 analysis jobs at each Tier-2
  site.} \label{fig:t2success}
\end{figure}

The STEP~09 jobs all read data from local disk at the sites, but did
not stage out any output, so the stageout elements of the analysis
model were not tested.  The majority of sites handled the STEP~09 jobs
perfectly, as indicated in Figure~\ref{fig:t2success}.  The overall
success rate for jobs was 80\%.  90\% of the job failures were due to
file read errors at the sites, which indicates a clear area that needs
improvement.  However, this indicates that the bulk of the problems
happened after jobs reached the sites, rather than during the grid
submission.  This would not have been true just a few years ago.

\section{Conclusions from STEP~09 and Outlook}

The STEP~09 exercise allowed us to focus on specific key areas of the
computing system in a multi-VO environment -- data transfers between
tiers, the use of tape systems at Tier~0 and Tier~1, and data analysis
at Tier~2.  Most of the Tier-1 sites showed good operational maturity.
Some may not yet have deployed all of the resources that will be
needed at LHC startup this fall, but there are no indications that
they will have any problem scaling up.  Not all Tier-1 sites attained
the goals of the tests; specific tests will be re-run after
improvements are made.  The tests of analysis activities at Tier~2
were largely positive.  Most sites were very successful, and CMS
easily demonstrated that it can use resources beyond the level pledged
by sites.  If anything, there are indicators that some resources could
be used more efficiently.

While STEP~09 gives us confidence that the CMS computing system
will work, there are still many challenges ahead of us.  For instance:
\begin{itemize}
\item The first run of the LHC will be longer than originally 
imagined.  What are the operational impacts?
\item If the LHC duty cycle is low at the start, there will be
pressure to increase the event rate at CMS, possibly to as high
as 2000~Hz from the nominal 300~Hz, and to overdrive the computing
systems.  Will it work?
\item Datasets will be divided into streams on the basis of triggers
  for custodial storage at the various Tier-1 sites.  This will allow
  re-processing to be prioritized by trigger type, but will the local
interests at each Tier-1 site by satisfied by the set of triggers it
stores?
\item Read errors were the leading problem in the Tier-2 analysis tests.
What can be done to make disk systems more reliable and maintainable?
\item The current system for remote stageout will not scale.  What will?
\item During a long run, will we be able to keep multiple copies of
RECO-level data available at Tier-2 sites?  If not, how will people
adjust?
\end{itemize}
We will learn a lot in the next year as we face up to these questions,
but we are confident that we are well-positioned to succeed.

\begin{acknowledgments}
  I thank my colleagues Ian Fisk, Oliver Gutsche and Frank
  W\"{u}rthwein for their assistance in preparing this presentation.
\end{acknowledgments}

\end{document}